# Digraph Decompositions and Monotonicity in Digraph Searching

Stephan Kreutzer and Sebastian Ordyniak

,
Oxford University Computing Laboratory,
University of Oxford
{kreutzer,ordyniak}@comlab.ox.ac.uk

October 29, 2018

**Abstract**

We consider monotonicity problems for graph searching games. Variants of these games – defined by the type of moves allowed for the players – have been found to be closely connected to graph decompositions and associated width measures such as path- or tree-width.

Of particular interest is the question whether these games are monotone, i.e. whether the cops can catch a robber without ever allowing the robber to reach positions that have been cleared before. The monotonicity problem for graph searching games has intensely been studied in the literature, but for two types of games the problem was left unresolved. These are the games on digraphs where the robber is invisible and lazy or visible and fast. In this paper, we solve the problems by giving examples showing that both types of games are non-monotone.

Graph searching games on digraphs are closely related to recent proposals for digraph decompositions generalising tree-width to directed graphs. These proposals have partly been motivated by attempts to develop a structure theory for digraphs similar to the graph minor theory developed by Robertson and Seymour for undirected graphs, and partly by the immense number of algorithmic results using tree-width of undirected graphs and the hope that part of this success might be reproducible on digraphs using a "directed tree-width". For problems such as disjoint paths and Hamiltonicity, it has indeed been shown that they are tractable on graphs of small directed tree-width. However, the number of such examples is still small.

We therefore explore the limits of the algorithmic applicability of digraph decompositions. In particular, we show that various natural candidates for problems that might benefit from digraphs having small "directed tree-width" remain NP-complete even on almost acyclic graphs.

# 1 Introduction

The seminal work of Robertson and Seymour in their graph minor project has focused much attention on graph decompositions and associated measures of graph connectivity such as tree- or path- width. Aside from the interest in graph structure theory, these notions have also proved fruitful in the development of algorithms. The tree-width of a graph is a measure of how tree-like the graph is and small tree-width allows for graph decompositions along which recursive algorithms can work. Many problems that are intractable in general can be solved efficiently on graphs of bounded tree-width. These include such classical NP-complete problems as finding a Hamiltonian-cycle in a graph or detecting if a graph is three-colourable. See [9, 8] and references therein for an introduction to tree-width.

Closely related to the theory of graph decompositions is the theory of graph searching games. In a graph searching game a number of searchers, or cops, tries to catch a fugitive, or robber, hiding in the graph. There are many variants of these games. The robber can hide on edges or vertices, he can be fast or lazy, he can be visible or not, the game can be played on undirected or directed graphs, and many more. Graph searching games are particularly interesting in relation to graph decompositions, as many width measures for graphs based on decompositions can also be described in terms of variants of Cops and Robber games. For instance, in 1993, Seymour and Thomas [12] showed that the tree-width of a graph equals the minimal number of cops required to catch a visible and fast robber (minus one). Dendris, Kirousis, and Thilikos [10] gave an analogous characterisation in terms of an invisible, lazy robber game. Other variants of Cops and Robber games have also been used to characterise the path-width of graphs and similar connectivity measures.

An important concept in the theory of graph searching games is monotonicity. A game is *monotone*, if whenever $k$ cops can catch a robber on a graph they can do so without allowing the robber to re-occupy vertices. In general, restricting the cops to monotone strategies may require additional cops to catch a robber. LaPaugh [20] gave a first proof of monotonicity for a Cops and Robber game. Since then, monotonicity has been intensely studied and a large number of monotonicity results have been established. See e.g. [20, 7, 10, 4, 13, 14, 21, 27] or the survey [2] and references therein.

Despite the considerable interest and the large number of results in this field, two cases have so far resisted any attempts to solve the monotonicity problem – the Cops and Robber game with a visible, fast robber and the game with an invisible, lazy robber, both played on digraphs. In this paper, we solve the problems by showing that both games are non-monotone.

**Digraph decompositions.** In recent years, attempts have been made to generalise the notion of tree-decompositions and their algorithmic applications to directed graphs. Clearly, we can define the tree-width of a directed graph as the tree-width of the undirected graph we get by forgetting the direction of edges, a process which leads to some loss of information. This loss may be significant, if the algorithmic problems we are interested in are inherently directed. A good example is the problem of detecting Hamiltonian cycles. While we know that this can be solved easily on graphs with small tree-width, there are directed graphs with very simple connectivity structure which have large tree-width. Therefore, several proposals have been made to extend the notions of tree-decompositions and tree-width to directed graphs (see [24, 17, 4, 6, 25, 16]). In particular, Reed [24] and Johnson, Robertson, Seymour, and Thomas [17] introduce the notion of *directed tree-width* and they show that Hamiltonicity can be solved for graphs of bounded directed tree-width in polynomial time.

Following this initial paper, several alternative definitions of directed graph decompositions have been proposed, with the aim of overcoming some shortcomings of the original definition. Obdržàlek [23] and Berwanger, Dawar, Hunter, and Kreutzer [5] introduce the notion of DAG-width and Hunter and Kreutzer [16] introduce the notion of Kelly-width. All three proposals are supported by algorithmic applications and various equivalent characterisations in terms of obstructions, elimination orderings, and, in particular, variants of Cops and Robber games on directed graphs. However, so far the algorithmic applications are restricted to few classes of problems, in particular the problem of finding disjoint paths, Hamiltonian-cycles, and similar linkage problems and certain problems in relation to combinatorial games (parity games) played on graphs that are motivated by the theory of computer-aided verification. Whereas the tree-width of undirected graphs has been employed to solve a huge number of problems on graphs of small tree-width, the algorithmic theory of directed graph decompositions is not nearly as rich.

It is an obvious question whether this is due to the fact that digraph decompositions are a relatively new field of research, where the fundamental machinery first needs to be developed, or whether this is due to a general limitation of this approach to algorithms on digraphs. In this paper we systematically explore the range of algorithmic applicability of digraph decompositions. For this, we look at typical NP-complete problems on graphs – as they can be found in [15] for instance – and identify those that are "suitable"for this approach, where by "suitable" we mean that the problems should be NP-complete in general but be tractable on acyclic graphs. The reason for the latter is that all digraph decompositions proposed so far measure in some way the similarity of a graph to being acyclic. In particular, acyclic graphs have small width in all of these measures. Hence, if a problem is already hard on acyclic digraphs, there is no point in studying the effect of digraph decompositions on this problem. We then identify representatives for the various types of "suitable" problems and ask whether they can be solved in polynomial time on graphs of small directed tree-width, Kelly- or DAG-width, or directed path-width.

The results we present in Section 4 show that the border for algorithmic applicability of digraph decompositions is rather tight. Essentially, as far as classical graph theoretical problems are concerned, disjoint paths and Hamiltonian-cycles can be detected efficiently on graphs of small directed tree-width, but all other problems we considered such as Minimum Equivalent Subgraph, Feedback Vertex Set (FVS), Feedback Arc Set, Graph Grundy Numbering, and several others are NP-complete even on graphs with a very low global connectivity and thus very low directed path or tree-width.

**Organisation.** The paper is organised as follows. In Section 2 we briefly recall basic notions from graph and game theory needed in the sequel. In Section 3 we give a formal description of Cops and Robber games and present the first main result of this paper, the non-monotonicity of the two types of games mentioned above. In Section 4 we explore the algorithmic boundaries of the digraph decompositions obtained so far by showing NP-completeness for a number of problems on digraphs with bounded "width". We conclude and state some open problems in Section 5.

## 2 Preliminaries

We use standard notation from graph theory as can be found in, e.g., [11]. All graphs and directed graphs in this work are finite and simple.

Let $G$ be a (directed) graph. We denote the vertexset of $G$ by $V(G)$ and the edgeset of $G$ by $E(G)$. For $X \subseteq V(G)$ we denote by $G[X]$ the subgraph of $G$ induced by $X$ and by $G \setminus X$ the subgraph of $G$ induced by $V(G) \setminus X$. Similarly for $Y \subseteq E(G)$ we set $G \setminus Y$ to be the subgraph of $G$ after deleting all edges in $Y$. A path in $G$ from a vertex $v_1$ to a vertex $v_n$ is a subgraph of $G$ with vertices $v_1, \cdots, v_n$ and edges $\{\{v_i, v_{i+1}\} \mid 1 \leq i < n\}$ if $G$ is undirected, respectively $\{(v_i, v_{i+1}) \mid 1 \leq i < n\}$ if $G$ is directed. For convenience we write $v \to u$ in $G$ if $G$ contains a path from $v$ to $u$. A cycle in $G$ is a path from $v_1$ to $v_n$ together with an edge $\{v_n, v_1\}$ if $G$ is undirected, and $(v_n, v_1)$ if $G$ is directed. We say $X \subseteq V(G)$ is connected, if for all pairs $x, y \in X$ there is a path from $x$ to $y$ in $G[X]$. A component in $G$ is a maximal connected vertexset in $G$.

Finally, for a set $X$ and $k \in \mathbb{N}$, we denote by $[X]^{\leq k}$ the set of all subsets of $X$ of cardinality $\leq k$.

## 3 Cops and robber games

Cops and Robber games are played by two players, that alternately place tokens on the vertices of a graph. Whereas the robber player has only one token and is merely able to move his token in a restricted way (depending on the variant of the game), the cop player can use an arbitrary amount of tokens and is free to move them anywhere on the graph. As the name suggests the objective of the cop player is to capture the token of the robber, i.e. to force the robber into a position where he is not able to move any more. Depending on the variant of the game – the variants differ in the abilities of both players – the minimum number of tokens needed by the cop to capture the robber defines a graph invariant.

More formally, let $D$ be a graph – either directed or undirected. A position in the game is a pair $(X, r)$, with $X \subseteq V(D)$ and $r \in V(D)$, and a play is a sequence of positions $((X_1, r_1), \cdots, (X_n, r_n))$, such that $X_1 = \emptyset$ and a move from one position to another is only allowed if the robber player is allowed to move from $r_i$ to $r_{i+1}$ with respect to $D \setminus (X_i \cap X_{i+1})$, $X_{i+1}$, and the variant of the game. For a play $((X_1, r_1), \cdots, (X_n, r_n))$ we define the robber-space as a sequence of vertexsets $(R_1, \cdots, R_n)$, with $R_1 = V(D)$ and $R_i = \{r \mid$ the robber can move from $r_{i-1}$ to $r\}$, for $i > 1$. The cop player wins, if there is a position satisfying $r_i \in X_i$, otherwise the robber player wins.

We are mainly interested in the type of *strategies* the players can employ. One can easily verify that strategies in these games only depend on the current position of the game, i.e. are deterministic and positional. Hence, strategies are functions assigning a new position for a player depending on the current position in the game. A strategy is *winning* for a player, if he wins all plays consistent with it, i.e. where all transitions from one position to another are consistent with it.

Let $D$ be a digraph, and $f$ a strategy for the cop. We define the *cop-width* of $f$, in terms $\mathrm{cw}(f)$, to be $\mathrm{cw}(f) = \max\{|f(X, r)| \mid X \subseteq V(D), r \in V(D)\}$, and the *cop-width* of $D$ to be $\mathrm{cw}(D) = \min\{cw(f) \mid$ f is winning on D $\}$. So the cop-width of a graph defines the graph invariant that we are interested in.

Before explaining the different variants of the game we introduce the concept of monotonicity. We say a play $((X_1, r_1), \cdots, (X_n, r_n))$ is cop-monotone, if the cop player never reoccupies a previously vacated vertex, i.e. there are no indices $1 \leq i < j \leq n$, such that $(X_i \setminus X_{i+1}) \cap (X_j \setminus X_{j-1}) \neq \emptyset$. We say a play is robber-monotone, if the corresponding robber-space never increases. A play is monotone, if it is both robber- and cop-monotone.

The notion of monotonicity directly applies to cop-strategies, so we say that a cop-strategy is robber-monotone, cop-monotone or just monotone, if all plays consistent with this strategy are. We denote by mon-$\mathrm{cw}(D) = \min\{cw(f) \mid$ f is monotone and winning on D $\}$ and say that a game is monotone if mon-$\mathrm{cw}(D) = \mathrm{cw}(D)$ for all graphs $D$, and non-monotone otherwise.

We are now ready to introduce the variants of the game. On undirected graphs a move from $(X, r)$ to $(X', r')$ is legal, if there exists a path from $r$ to $r'$ in $D \setminus (X \cap X')$, i.e. the robber is allowed to move along cop free paths. The variant of the game that places no other restriction on the robber is called *dynamic* as the robber is allowed to move in every move of the game, except when he is captured. Contrary to that is the so called *inert* variant, where the robber is only able to move when the cop player is going to occupy his current position, i.e. a move from $(X, r)$ to $(X', r')$ is legal if $r \in X'$. Furthermore there is also a variant of the game where the cops are unable to see the robber, which is called *invisible*. The normal version, i.e. where the cops can see the robber is called *visible*.

Combining these, one obtains four variants of the game, of which only three are considered in literature, namely: visible and dynamic (vis), invisible and dynamic (invis), and invisible and inert (inert). On undirected graphs all these variants are monotone and satisfy:

1. vis-$\mathrm{cw}(D) =$ inert-$\mathrm{cw}(D) = \mathrm{tw}(D) + 1$, for every graph $D$, where $\mathrm{tw}(D)$ denotes the tree-width of $D$ ( see [12] and [10] ).

2. invis-cw($D$) = pw($D$) + 1, for every graph $D$, where pw($D$) denotes the path-width of $D$ ( see [7] ).

Depending on how one translates the notion of an undirected path to the directed setting, i.e. whether one regards it as one directed path from source to destination or as two directed paths ,one in each direction, there are two natural variants of this game on directed graphs. We refer to the first variant, i.e. where the robber is allowed to move along (cop-free) directed paths, as reachability variant (reach), and to the second one, i.e. where the robber is only allowed to move when there exist a path in each direction, as strong connected component (scc) variant, since in this case the robber is only allowed to move in strongly connected components.

Combining these two main versions of the game with the variants discussed for the undirected setting one retrieves a number of interesting games on directed graphs of which the following have been discussed in literature so far: strong connected component, visible and dynamic (scc-vis); reachability, visible and dynamic (reach-vis); reachability, invisible and dynamic (reach-invis); and reachability, invisible and inert (reach-inert). We briefly relate these games to the corresponding digraph decompositions and recall what is known about monotonicity.

**scc, visible, and dynamic:** This variant is closely related to directed tree-width as it is known that scc-vis-cw($D$) $-1 \leq dtw(D) \leq 3 \cdot$ scc-vis-cw($D$) $+5$, for every digraph $D$ with directed tree-width $dtw(D)$ (see [18]). It has been shown to be neither robber- nor cop-monotone [1, 17]. However, although not explicitly stated, [17] gives an upper bound for the monotonicity costs with respect to robber-monotonicity. It remains an interesting open question whether this holds for the cop-monotone variant as well.

**reachability, invisible and dynamic:** This variant defines directed path-width and has been shown to be monotone in [4].

**reachability, visible and dynamic:** The monotone version of this variant defines DAG-width [5]. We therefore refer to these games as *DAG-games*.

**reachability, invisible and inert:** The monotone version of this variant defines Kelly-width [16]. We therefore refer to these games as *Kelly-games*.

We are now ready to state our main results of this section, proving that DAG- and Kelly-Games are non-monotone.

### 3.1 Non-Monotonicity of DAG-Games

**Theorem 3.1.** *For every $p \geq 2$ there exists a digraph $D_p$ with* mon-dag-cw($D_p$) $= 4p - 2$ *and* dag-cw($D_p$) $= 3p - 1$.

*Proof.* A schematic overview of $D_p$ is given in Figure 1. The graph consists of three main parts with $2p - 1$ vertices each. $C_0$ and $C_2$ are cliques on $2p - 1$ vertices, $C_1^2$ is a clique on $p - 1$ vertices and $C_1^1$ forms an independent set having $p$ vertices. A directed edge between two parts $A$ and $B$ means that there are edges from every vertex in $A$ to every vertex in $B$. Undirected edges mean that there are edges between $A$ and $B$ in both direction.

It is easy to see that dag-cw($D_p$) $\geq 3p - 1$ since the vertices in $C_0 \cup C_1^2$ together with a vertex of $C_1^1$ form a clique of size $3p - 1$. To show that dag-cw($D_p$) $\leq 3p - 1$ consider the following strategy for $3p - 1$ cops on $D_p$. In the first move the cops occupy $C_0 \cup C_1^1$. If the robber plays to $C_2$ the cops capture him by playing on $C_1^1 \cup C_2$. Otherwise, if the robber plays to $C_1^2$ the cops move to $C_0 \cup C_1^2$.

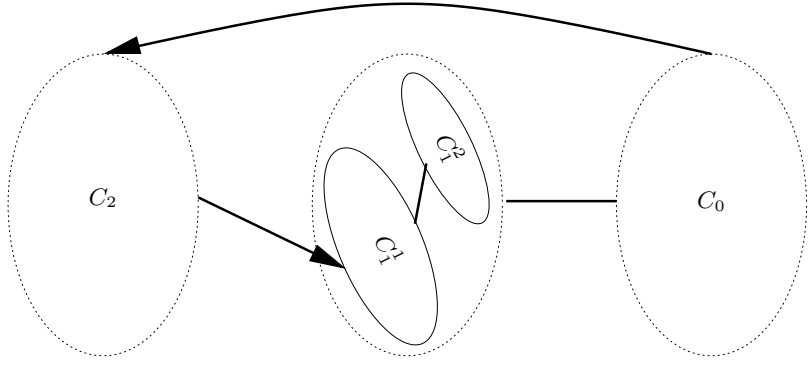

Figure 1: The graph $D_p$ with dag-cw$(D_p) \neq$ mon-dag-cw$(D_p)$.

Now the robber has to be on a vertex $v \in C_1^1$. Since the vertices in $C_1^1$ form an independent set the robber is now captured by playing to $\{v\} \cup C_1^2 \cup C_0$.

It remains to show that mon-dag-cw$(D_p) = 4p - 2$. It is easy to see that $4p - 2$ cops can capture the robber on $D_p$ by playing $C_0 \cup C_2$ and then $C_0 \cup C_1$. To show that mon-dag-cw$(D_p) \geq 4p - 2$ we give a strategy for the robber against $4p - 3$ cops playing monotonously on $D_p$.

First the robber stays in $C_0$ until the cops occupy all vertices of $C_0$. There are two cases to consider.

1. The cops occupy ( at least ) $C_0 \cup C_1^1$. In this case there is a vertex $v \in C_1^2$ which is not occupied by a cop and which the robber can reach from his current position in $C_0$. Since every $v \in C_1^2$ has an edge to every other vertex in $C_0 \cup C_1$ the cop cannot capture the robber monotonously with less than $4p - 2$ cops.

2. The cops occupy ( at least ) $C_0$ and there is at least one vertex in $C_1^1$ which is not occupied by a cop. Then there exists a vertex $v \in C_2$ which is not occupied by a cop and which the robber can reach from his current position in $C_0$. Since from every vertex in $C_2$ there is a path to every other vertex in the graph (as long as there is at least one vertex in $C_1^1$ not occupied by a cop) the robber can stay in $C_2$ until the cops occupy all vertices in $C_1^1$. And if they do the robber can move to a vertex in $C_1^2$ and play as in the first case.

□

## 3.2 Non-Monotonicity of Kelly-Games

We now consider Kelly-games. Recall that in a Kelly-game, the robber is invisible. Hence, a strategy for the cop must be independent of the current position of the robber. We can therefor represent a cop-strategy in a digraph $D$ by a sequence $(v_1, \ldots, v_{|D|})$ of vertices in the order in which they are visited by the cops.

**Theorem 3.2.** *For every $p \geq 2$ there exists a digraph $D_p$ with* mon-kelly-cw$(D_p) = 7p$ *and* kelly-cw$(D_p) = 6p$.

*Proof.* A schematic overview of $D_p$ is given in Figure 2. The graph consists of five cliques with $|C_0| = p$, $|C_2| = |C_1| = |X_1| = 2p$, $|X_2| = 3p$. An edge between two parts $A$ and $B$ means that there are edges from every vertex in $A$ to every vertex in $B$, where again an undirected edge between $A$ and $B$ means that there are edges in $D_p$ in both directions.

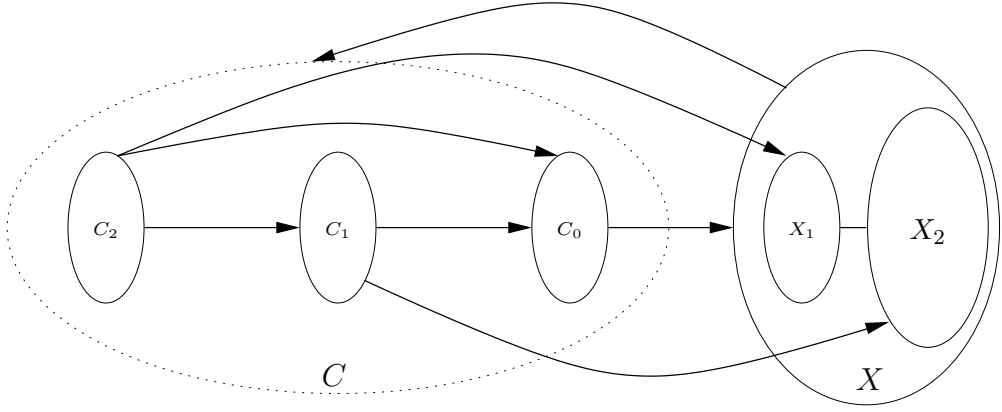

Figure 2: The graph $D_p$ with kelly-cw$(D_p) \neq$ mon-kelly-cw$(D_p)$.

The following strategies for the cop show that mon-kelly-cw$(D_p) \leq 7p$ and kelly-cw$(D_p) \leq 6p$. For the monotone game we use the strategy $(X \cup C_0, X_2 \cup C_0 \cup C_1, X_1 \cup C)$, i.e. the cops first occupy all of $X$ and $C_0$, then proceed to $X_2 \cup C_0 \cup C_1$, and finally move to $X_1 \cup C$. For the non-monotone case we use $(X \cup C_0, X_2 \cup C_0 \cup C_1, X_1 \cup C_1, X_1 \cup C_1 \cup C_2, X, X \cup C_0)$.

To see that kelly-cw$(D_p) \geq 6p$ note that $C_0 \cup X$ is a clique of size $6p$. It remains to show that mon-kelly-cw$(D_p) \geq 7p$. Suppose mon-kelly-cw$(D_p) < 7p$ and let $S = (v_1, \cdots, v_{|V(D_p)|})$ be a cop-strategy witnessing this. For each part $Y \in \{C_0, C_1, C_2, X_1, X_2, C, X\}$ of $D_p$ let $I(Y)$ be the greatest index of a vertex in $Y$, i.e. $v_{I(Y)}$ is the last vertex of $Y$ which is searched by $S$. Then the following statements hold:

1. $I(X) < I(C_1)$ and $I(X) < I(C_2)$. For the sake of contradiction, suppose $I(X) > I(C_1)$ and let $v = v_{I(X)}$. Hence, when the cops clear $v$, they have already cleared all vertices in $X$ other than $v$ and all vertices in $C_1$. As $v$ has edges to every other vertex in $C_1 \cup X$, the cops need to occupy all of $(C_1 \cup X) \setminus \{v\}$ before they can place a token on $v$. But this requires $7p$ cops.

   The case of $I(X) < I(C_2)$ is analogous.

2. $I(C_0) < I(C_1)$. Again, assume the contrary, i.e. $I(C_0) > I(C_1)$. Hence, when clearing $v_{I(C_1)}$ there is a free vertex $v \in C_0$ through which the robber can reach all of $X$. As $I(X) < I(C_1)$, the cops needs to occupy at least $(X \cup C_1) \setminus \{v_{I(C_1)}\}$ before clearing $v_{I(C_1)}$, which yields the contradiction.

3. $I(C_1) < I(C_2)$. With a similar reasoning as before we obtain that otherwise the cops have to occupy $X \cup C_2$ when searching $v_{I(C_2)}$, using $7p$ cops.

The statements (1)-(3) imply $I(X) < I(C_0) < I(C_1) < I(C_2)$ but now the cop needs to occupy $|C_2 \cup C_1 \cup C_0 \cup X_1| = 7p$ vertices in order to search $v_{I(C_2)}$. So $S$ uses at least $7p$ cops. □

## 4 Limits of Algorithmic Applications

In [17] it has been shown that the $k$-disjoint path problem and related problems are solvable in polynomial time on graphs of bounded directed tree-width. However, up to now only few other problems are known to be solvable with the help of digraph decompositions, a further example being parity games, which are tractable on graphs of bounded DAG- and Kelly-width [5, 16]. As directed tree-width is the most general of these width-measures, tractability results for directed tree-width directly extend to all other measures. The converse is not true, for example it is not known whether parity games are tractable on graphs of bounded directed tree-width.

In this section we explore the algorithmic boundaries of the digraph measures introduced so far. In our analysis we focus on NP-complete problems that are explicitly directed. All analysed problems are solvable in polynomial time on digraphs whose underlying undirected graph has bounded tree-width – but as mentioned in the introduction, tree-width is not a good measure for the global connectivity of a digraph. Furthermore, we discard problems that are not tractable on acyclic graphs, as all measures defined so far are bounded on acyclic graphs. As representatives for various types of the remaining problems, we have considered the following problems: Minimum Equivalent Subgraph, Directed Feedback Vertex / Arc Set, Graph Grundy Numbering, and Kernel.

It turned out that all of these problems remain NP-complete even on digraphs that have very low global connectivity, i.e. digraphs that can be decomposed into strong components of constant size just by removing a small number of vertices. In particular, these graphs have low width with respect to all digraph decompositions defined so far, i.e. small directed path width, small DAG-, Kelly-, and directed tree-width, small Entanglement and D-width. For notational convenience, we state the proofs in terms of DAG-width, which as already stated in Section 3 is equal to the number of cops needed to catch the robber in the reachibility, monotone, visible and dynamic cops and robber game.

### 4.1 Minimum Equivalent Subgraph

The *Minimum Equivalent Subgraph (MES)*-problem is the problem to compute in a given digraph $D$ an edge-minimal subgraph $D' \subseteq D$ that preserves reachability in $D$.

**Definition 4.1.** *Let $D$ be a digraph and $k \in \mathbb{N}$. MES is the problem to decide, if there is a set $E' \subseteq E(D)$ with $|E'| \leq k$, such that the digraph $D' = (V(D), E')$ contains a path between two vertices if, and only if, such a path exists in $D$, i.e. $D$ and $D'$ have the same transitive closure.*

MES is NP-complete for arbitrary digraphs (see [15]), but is known to be solvable in polynomial time for acyclic and undirected graphs. In [22] it is also shown that it suffices to consider MES on connected digraphs. There MES is equivalent to a generalisation of the directed hamilton cycle problem, the so-called round-trip-problem, in which vertices can be used more than ones. This is particularly interesting because the directed hamilton cycle problem is a special case of the $k$-linkage problem, which can be solved in polynomial time on digraphs of bounded DAG-width.

**Definition 4.2.** *Let $D$ be a connected digraph. A round-trip $R = (v_1, \cdots, v_k, v_1)$ is a sequence of $k + 1$ vertices of $D$, such that $(v_i, v_{i+1}) \in E(D)$ and $R$ visits every vertex of $D$ at least once. The size of $R$ equals $k + 1$.*

**Lemma 4.3.** *[22] Let $D$ be a connected digraph and $k$ a natural number. Then $D$ has a MES of size less than $k$ if, and only if, $D$ has a round-trip of size less than $k$.*

The NP-completeness of MES for digraphs of DAG-width less than four follows from a reduction of 3-SAT to the problem of finding a minimum round-trip in a connected digraph of DAG-width less than four as follows:

**Theorem 4.4.** *The MES-problem is NP-complete on directed graphs of DAG-width less than four.*

*Proof.* The proof reduces 3-SAT to round-trip. Let $F$ be a 3-SAT-Formula with variables $x_1, \cdots, x_n$ and clauses $C_1, \cdots, C_m$. From $F$ we construct a digraph $D$ satisfying:

**(A)** $F$ is satisfiable if, and only if, $D$ has a round-trip of size $|V(D)| + m$.

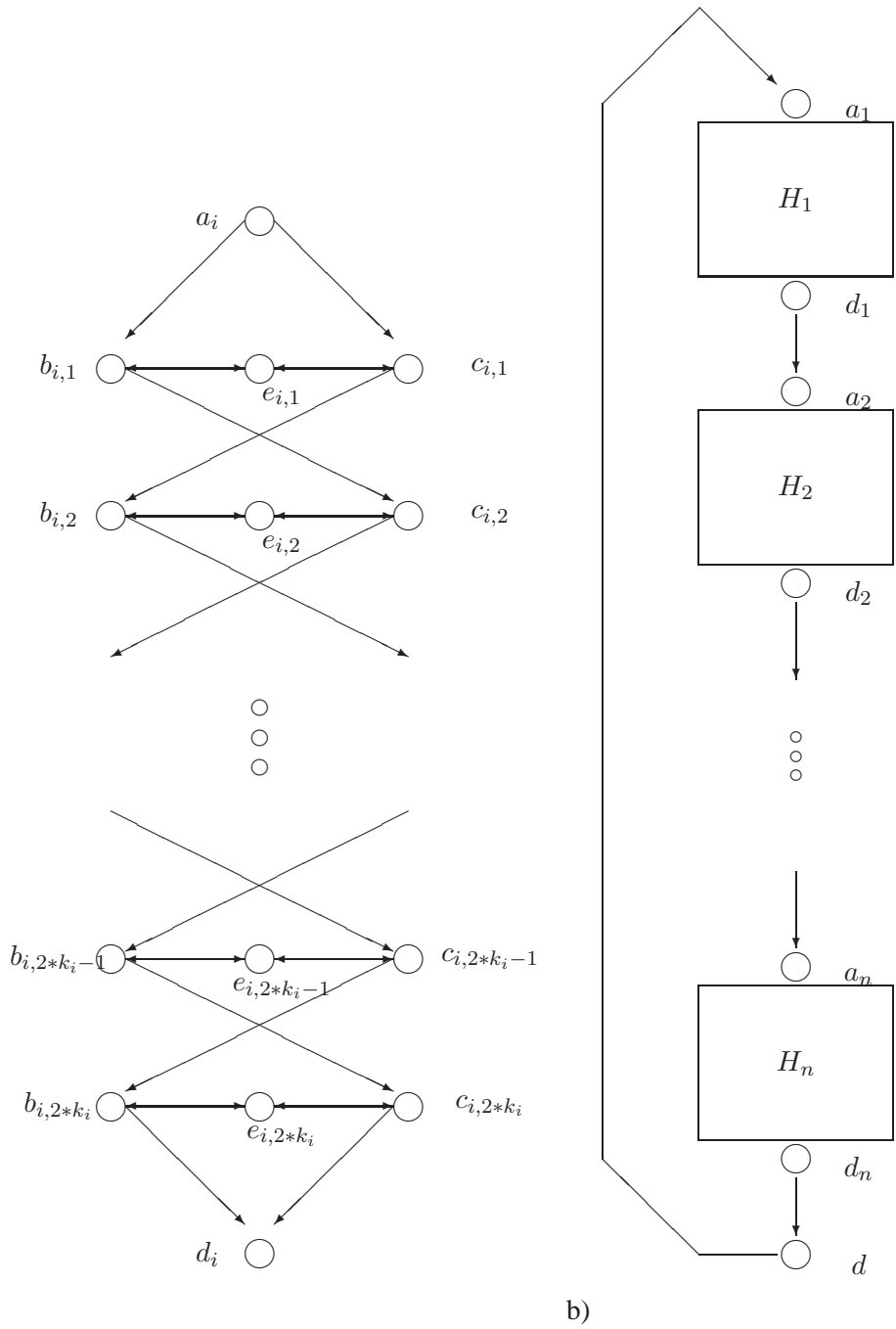

Figure 3: a) The graph $H_i$ for the reduction of 3-SAT to MES in Theorem 4.4. b) The connections between the graphs $H_1, \cdots, H_n$ and the vertex $d$.

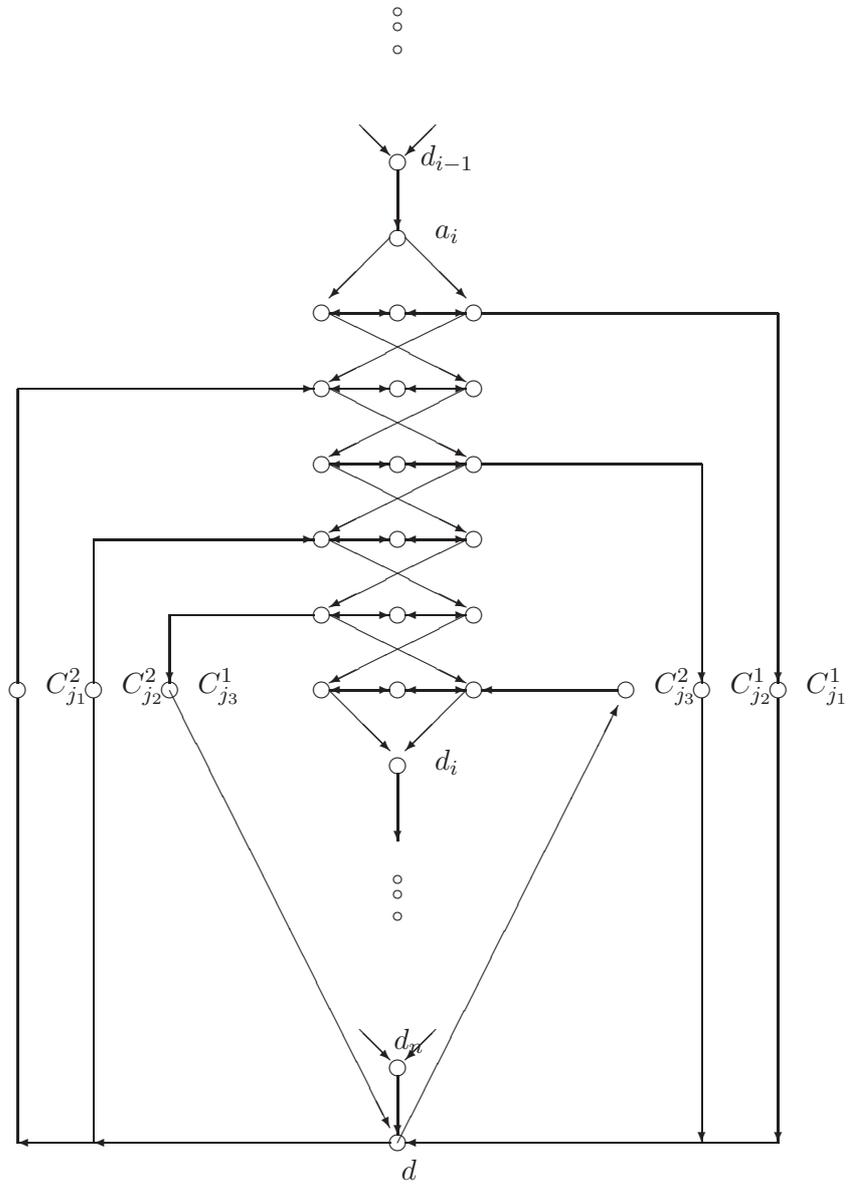

Figure 4: An example for the reduction of 3-SAT to round-trip in Theorem 4.4. In this example $x_i$ is positive in $C_{j_1}$, $C_{j_2}$ and negative in $C_{j_3}$.

**(B)** $D$ is constructable in polynomial time, given $F$.

**(C)** $D$ has DAG-width less than four.

We construct $D$ as follows:

- For each variable $x_i$ in $F$ add the digraph $H_i$, which is shown in Figure 3 a), to $D$.
- Add a vertex $d$ to $D$.
- Connect $d$ and the graphs $H_i$ as illustrated in Figure 3 b), i.e. add the edges $(d_i, a_{i+1})$, $(d_n, d)$ and $(d, a_1)$ to $D$.
- For each clause $C_j$ add the vertices $C_j^1$ and $C_j^2$ together with the edges $(C_j^1, d)$ and $(d, C_j^2)$ to $D$.
- For each occurrence of $x_i$, respectively $\overline{x}_i$ in a clause $C_j$ add the edges $(b_{i,2*l-1}, C_j^1)$ and $(C_j^2, c_{i,2*l})$, respectively $(c_{i,2*l-1}, C_j^1)$ and $(C_j^2, b_{i,2*l})$ to $D$, where $l$ is the smallest integer, such that neither $b_{i,2*l-1}$ nor $c_{i,2*l-1}$ have been used by a clause before.

An example for $D$ is shown in Figure 4. It remains to show that $D$ actually satisfies (A)-(C).

**(A)** $\implies$ Let $\beta$ be a satisfying assignment for $F$. We have to show that $D$ contains a round-trip $R$ of size $|V(D)| + m$. $R$ uses the following edges:

- $(d_i, a_{i+1})$, $(d_n, d)$ and $(d, a_1)$.
- For each $H_i$, such that $\beta(x_i) = $ true, respectively $\beta(x_i) = $ false, $R$ uses the edges $(a_i, b_{i,1})$, $(b_{i,1}, e_{i,1})$ and $(e_{i,1}, c_{i,1})$, respectively $(a_i, c_{i,1})$, $(c_{i,1}, e_{i,1})$ and $(e_{i,1}, b_{i,1})$.
- If $\beta(x_i) = $ true, respectively $\beta(x_i) = $ false and $c_{i,l}$, respectively $b_{i,l}$ has an edge to a clause $C_j$, such that neither $C_j^1$ nor $C_j^2$ are already contained in $R$, use the edges $(c_{i,l}, C_j^1)$, $(C_j^1, d)$, $(d, C_j^2)$ and $(C_j^2, b_{i,l+1})$, respectively $(b_{i,l}, C_j^1)$, $(C_j^1, d)$, $(d, C_j^2)$ and $(C_j^2, c_{i,l+1})$. If not $R$ uses the edge $(c_{i,l}, b_{i,l+1})$, respectively $(b_{i,l}, c_{i,l+1})$.

It is easy to see that $R$ is a round-trip for $D$, using every vertex except $d$ at most once and every vertex in $H_1, \cdots, H_n$ exactly once. As $\beta$ is a satisfying assignment for $F$ every vertex $C_j^1$, $C_j^2$ is used at least once and $d$ is used exactly $m+1$-times. So $R$ has size $|V(D)| + m$.

$\impliedby$ Now suppose we are given a round-trip $R$ on $D$ of size $|V(D)| + m$. We have to show that there exist a satisfying assignment for $F$. We show this by a series of claims:

a) In every round-trip $d$ has exactly $m+1$ predecessors.

b) In every round-trip $d$ has exactly $m+1$ successors.

c) All vertices in $R$ except $d$ have in- and out-degree one and $d$ has in- and out-degree $m+1$.

d) $R$ contains exactly one of $(b_{i,j}, e_{i,j})$ and $(e_{i,j}, b_{i,j})$. The same holds for $(c_{i,j}, e_{i,j})$ and $(e_{i,j}, c_{i,j})$.

e) $R$ contains either $(b_{i,j}, e_{i,j})$ and $(e_{i,j}, c_{i,j})$ or $(c_{i,j}, e_{i,j})$ and $(e_{i,j}, b_{i,j})$ for every $i$ and $j$.

f) $R$ contains either all edges $(b_{i,j}, e_{i,j})$ and $(e_{i,j}, c_{i,j})$ or all edges $(c_{i,j}, e_{i,j})$ and $(e_{i,j}, b_{i,j})$, for every $i$.

g) If $R$ contains an edge $(b_{i,j_1}, C^1_{l_1})$, than $R$ does not contain an edge $(c_{i,j_2}, C^1_{l_2})$, for every $i$.

h) The assignment $\beta$, with $\beta(x_i) =$ true, if $R$ contains edges of the form $(b_{i,j}, C^1_l)$ and $\beta(x_i) =$ false otherwise is a satisfying assignment for $F$.

Proof of (a)-(h):

a) This follows from the fact that $d$ is the only successor of it's $m+1$ predecessors in $D$.

b) This follows from the fact that $d$ is the only predecessor of it's $m+1$ successors in $D$.

c) This follows from (a) and (b) together with the fact that the size of $R$ is $|V(D)| + m$.

d) Suppose $R$ contains concurrently $(b_{i,j}, e_{i,j})$ and $(e_{i,j}, b_{i,j})$. Then it follows from (c) that $b_{i,j}$ and $e_{i,j}$ are isolated in $R$, a contradiction.

e) This follows from (d) and the fact that $b_{i,j}$ and $c_{i,j}$ are the only neighbours of $e_{i,j}$.

f) We show this by induction on $j$. For $j = 1$ this follows from (e). W.l.o.g. we can assume that $R$ contains $(b_{i,j}, e_{i,j})$ and $(e_{i,j}, c_{i,j})$. We have to show that $R$ also contains $(b_{i,j+1}, e_{i,j+1})$ and $(e_{i,j+1}, c_{i,j+1})$. As $(e_{i,j}, c_{i,j})$ is contained in $R$, $c_{i,j}$ has either $C^1_l$ or $b_{i,j+1}$ as successor in $R$. We therefore distinguish two cases:

  (a) $R$ contains $(c_{i,j}, C^1_l)$. Then the only predecessors of $c_{i,j+1}$ in $D$ are $b_{i,j}$ and $e_{i,j+1}$. As $b_{i,j}$ already has a successor in $R$, $R$ has to contain $(e_{i,j+1}, c_{i,j+1})$.
  
  (b) $R$ contains $(c_{i,j}, b_{i,j+1})$. In this case $b_{i,j+1}$ cannot have another predecessor in $R$, thus $R$ cannot contain $(e_{i,j+1}, b_{i,j+1})$.

g) Because of (f) for every $i$ either all vertices $b_{i,j}$ are succeeded by $e_{i,j}$ or all vertices $c_{i,j}$ are succeeded by $e_{i,j}$ in $R$. So $R$ contains either only edges of the form $(c_{i,j}, C^1_l)$ or only edges of the form $(b_{i,j}, C^1_l)$.

h) This follows from (g) and the fact that $R$ has to contain all vertices of $D$, in particular $C^l_j$.

**(B)** This follows from the construction of $D$.

**(C)** The following defines a monotone winning strategy $f$ for the cop-player on $D$ using less than 4 cops:

1) $f(\emptyset, r) = \{d, C^2_1\}$
2) $f(\{d, C^2_i\}, r) = \{d, C^2_{i+1}\}$ where $1 \leq i < m$.
3) $f(\{d, C^2_m\}, r) = \{d, a_1\}$
4) $f(\{d, a_i\}, r) = \{d, b_{i,1}, e_{i,1}\}$ where $1 \leq i \leq n$.
5) $f(\{d, b_{i,j}, e_{i,j}\}, r) = \{d, e_{i,j}, c_{i,j}\}$ where $1 \leq i \leq n$ and $1 \leq j \leq 2 \cdot k_i$.
6) $f(\{d, e_{i,j}, c_{i,j}\}, r) = \{d, a_{i,j+1}, e_{i,j+1}\}$ where $1 \leq i \leq n$ and $1 \leq j \leq 2 \cdot k_i$.
7) $f(\{d, e_{i,2 \cdot k_i}, c_{i,2 \cdot k_i}\}, r) = \{d, d_i\}$ where $1 \leq i \leq n$.
8) $f(\{d, d_i\}, r) = \{d, a_{i+1}\}$ where $1 \leq i < n$.
9) $f(\{d, d_n\}, r) = \{d, C^1_1\}$.
10) $f(\{d, C^1_i\} r) = \{d, C^1_{i+1}\}$ where $1 \leq i < m$.

□

## 4.2 Feedback Vertex Set / Feedback Arc Set

The *Feedback Vertex/Arc Set (FVS/FAS)*-problem is the problem to find a minimum set of vertices (edges) in a digraph $D$, whose removal leaves $D$ acyclic. Both problems are known to be NP-complete on arbitrary digraphs (see [19]). Trivially both problems become efficiently solvable on acyclic graphs. FVS is the only problem we present here that is NP-complete on undirected graphs as well.

We prove the NP-completeness of FVS/FAS on digraphs of DAG-width four by reducing it to a special variant of 3-SAT namely 3-SAT-2, which we introduce now.

**Definition 4.5.** *3-SAT-2 is the variant of 3-SAT, so that every literal is used in at most two clauses.*

3-SAT-2 is NP-complete. Next we need a simple lemma that helps us with the actual reduction.

**Lemma 4.6.** *Let $D$ be the complete bipartite graph with two vertices in each part. Then every FVS of $D$ contains all vertices of one part. Furthermore the vertices of one part form a FVS.*

*Proof.* Suppose for a contradiction that $V'$ is a FVS of $D$ and $a, b \notin V'$ are two vertices in $V'$ not in the same part of $D$. Then $(a, b), (b, a)$ is a circle in $D$ which is not covered by $V'$. Now suppose $V'$ is a part of $D$. Then $D - V'$ is an independent set and acyclic. □

We are now ready to prove the main theorem of this section.

**Theorem 4.7.** *FVS is NP-complete on graphs of DAG-width at most four.*

*Proof.* We reduce 3-SAT-2 to FVS on graphs of DAG-width at most 4. For every 3-SAT-2 formula $F$ with variables $x_1, \cdots, x_n$ and clauses $C_1, \cdots, C_m$ we construct a graph $D$ satisfying:

**(A)** $D$ has a FVS of size $2n$ if, and only if $F$ is satisfiable.

**(B)** $D$ is constructable in polynomial time given $F$.

**(C)** $D$ has DAG-width at most four.

$D$ is constructed by the following steps:

- Add a vertex $v$ to $D$.

- For each variable $x_i$ add to $D$ the complete bipartite graph $H_i$ with partitions $\{x_i^1, x_i^2\}$ and $\{\overline{x}_i^1, \overline{x}_i^2\}$.

- For each clause $C_j$ with literals $l_1, \cdots, l_k$, which are ordered corresponding to the index of their variables, add a circle $c_1, \cdots, c_k, v, c_1$ to $D$, such that :

  a) If $l_h$ equals $x_i$, respectively $\overline{x}_i$, then $c_h$ is one of $x_i^1, x_i^2$, respectively $\overline{x}_i^1, \overline{x}_i^2$.
  
  b) The vertex used by $l_h$ is not used by any other clause. This is always possible as each literal is contained in at most two clauses.

We now show that $D$ satisfies (A)-(C).

**(A)** $\implies$ Let $\beta$ be a satisfying assignment for $F$. Now define $V' \subseteq V(D)$, such that $x_i^1, x_i^2 \in V'$, if $\beta(x_i) = $ true and $\overline{x}_i^1, \overline{x}_i^2 \in V'$ otherwise. So $|V'| \leq 2n$ it remains to show that $V'$ is a Feedback Vertex Set for $D$. As $H_i \setminus V'$ is an independent set, no circle can contain an edge of an $H_i$. It follows from the construction of $D$ that every such circle corresponds to a unique clause. As $\beta$ is a satisfying assignment for $F$, $V'$ contains at least one vertex for each such a clause. So $V'$ is a FVS for $D$.

$\Longleftarrow$ Let $V'$ be a FVS for $D$ of size $2n$. Because of Lemma 4.6 $V'$ contains either $x_i^1, x_i^2$ or $\overline{x}_i^1, \overline{x}_i^2$ for each $1 \leq i \leq n$. As $|V'| \leq 2n$ $V'$ contains no other vertices. Now define $\beta(x_i) =$ true if $x_i^1, x_i^2 \in V'$ and $\beta(x_i) =$ false otherwise. As $V'$ is a FVS and $v \notin V'$, $V'$ contains at least one vertex of every clause. So $\beta$ is a satisfying assignment for $F$.

This follows easily from the construction.

To show that $D$ has DAG-width at most four, we give a monotone winning strategy for four cops on $D$:

1. $f(\emptyset, r) = \{v, x_1^1, x_1^2, \overline{x}_1^1\}$
2. $f(\{v, x_i^1, x_i^2, \overline{x}_i^1\}, r) = \{v, x_i^1, x_i^2, \overline{x}_i^2\}$ where $1 \leq i \leq n$
3. $f(\{v, x_i^1, x_i^2, \overline{x}_i^2\}, r) = \{v, x_{i+1}^1, x_{i+1}^2, \overline{x}_{i+1}^1\}$ where $1 \leq i < n$

□

To show that FAS is NP-complete on digraphs of bounded DAG-width as well, we use a simple reduction to FVS, as follows:

**Definition 4.8.** *Let $D$ be a digraph. Then $K(D)$ is the digraph obtained from $D$ after replacing every vertex $v \in V(D)$ with two vertices $v_1, v_2$ and an edge $(v_1, v_2)$ and every edge $(v, w) \in D$ with an edge $(v_2, w_1)$.*

In [3] it is shown that $D$ has a FVS of size $k$ if, and only if $K(D)$ has a FAS of size $k$. Thus using Lemma 4.7, it only remains to show that the DAG-width of $D$ equals the DAG-width of $K(D)$, for every digraph $D$.

**Lemma 4.9.** *Let $k \geq 2$ and $D$ be a digraph. Then the DAG-width of $K(D)$ is at most the DAG-width of $D$.*

*Proof.* Let $f$ be a monotone winning strategy for $k$ cops on $D$ and let $r \in V(D)$, $X \subseteq V(D)$. W.l.o.g. we can assume that $f$ either places or removes one cop at a time. If $f$ places a cop on a vertex $v$ then $f'$ copies this move by placing a cop on $v_2$. If $f$ removes a cop from a vertex $v$ and the robber does not occupy $v_1$, $f'$ just removes a cop from $v_2$. If the robber occupies $v_1$, when $f$ is to remove a cop from $v$ then $f'$ first removes all cops currently occupied except $v_2$ and after that, places a cop on $v_1$. Now the robber is captured on $v_2$. It is easy to see that $f'$ is a monotone winning strategy for $\max\{2, k\} = k$ cops on $K(D)$. □

Combining the previous lemma with theorem 4.7 we get:

**Theorem 4.10.** *FAS is NP-complete on graphs of DAG-width at most four.*

### 4.3 Graph Grundy Numbering and Kernel

**Definition 4.11.** Graph Grundy Numbering *is the problem to decide for a digraph $D$ if there exists a function $f : V(D) \to \mathbb{N}$, such that for all $v \in V(D)$, $f(v)$ is the smallest natural number not contained in $\{f(u) : u \in V(D), (v, u) \in E(D)\}$.*

**Definition 4.12.** Kernel *is the problem to decide in a digraph $D$ if there exists $V' \subseteq V(D)$, such that*

1. *there is no edge between two vertices in $V'$, i.e. $V'$ is an independent set.*

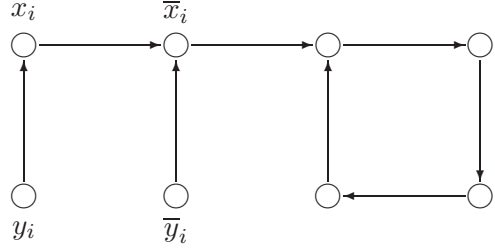

Figure 5: The graph $H_i$ for a variable $x_i$ used in the proof of Theorem 4.13.

2. *for every $v \in V(D) \setminus V'$ there exists a $u \in V'$ with $(v, u) \in E(D)$.*

Observe, that on undirected graphs the maximisation version of Kernel is the *Independent Set*-problem, whereas the minimisation version of Graph Grundy Numbering equals Vertex-Colouring. On digraphs however even the existential versions of both problems are known to be NP-complete [26], but are trivially solvable on acyclic graphs. A simple example of a digraph that neither has a Graph Grundy Numbering nor a Kernel is the directed cycle with three vertices. We are now ready to prove the NP-completeness for Graph Grundy Numbering on digraphs of DAG-width two.

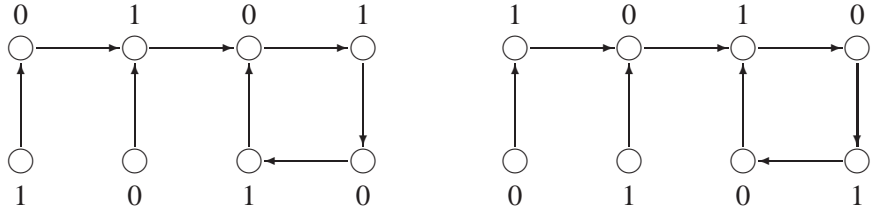

Figure 6: The two possible colourings for the graph in Figure 5.

**Theorem 4.13.** *Graph Grundy Numbering is NP-complete for digraphs of DAG-Width at most two.*

*Proof.* As the proof uses the reduction of 3-SAT to Graph Grundy Numbering given in [26], we only show that the graph used in [26] has DAG-width at most two. To do this we first take a look at what the graph looks like.

Let $F$ be a 3-SAT-Formula with variables $x_1, \cdots, x_n$ and clauses $C_1, \cdots, C_m$, then the digraph $D$ used in the reduction from 3-SAT to Graph Grundy Numbering is constructed as follows:

- For each variable $x_i$ we add the graph $H_i$, which is given in Figure 5, to $D$.

- For each clause $C_j$ we add the vertices $A_j, B_j, C_j$ and edges $(A_j, B_j), (B_j, C_j), (C_j, A_j)$ to $D$.

- For each occurrence of a variable $x_i$, respectively $\overline{x}_i$ in a clause $C_j$ we add to $D$ the edges $(C_j, y_i)$, respectively $(C_j, \overline{y}_i)$.

We now show that $D$ has DAG-width at most two. This is done by giving a description of a monotone winning strategy for two cops on $D$:

1. The robber starts on a vertex of $H_i$. In this case the robber cannot leave $H_i$ as there is no edge from a vertex in $H_i$ to a vertex in $D \setminus H_i$. As $H_i$ can be made acyclic by removing one vertex, $H_i$ can be searched monotonously by two cops.

2. The robber starts on one of the vertices $A_j, B_j, C_j$. As $D[\{A_j, B_j, C_j\}]$ is a circle and the robber is not able to reach a vertex $A_p, B_p, C_p$ for $p \neq j$ the two cops can push the robber to a $H_i$, where he can be captured as shown in case one.

□

**Theorem 4.14.** *Kernel is NP-complete for digraphs of DAG-width two.*

*Proof.* Since the prove uses the same graph as in theorem 4.13 and the reduction is given in [26] the result follows. □

## 5 Conclusion and Open Problems

In this paper we considered graph searching games on directed graphs and established non-monotonicity for two important types of games. Our examples show that the monotonicity costs for these games can not be bound by an additive term, i.e. for any $k$ there are digraphs where at least $k$ additional cops are required to catch a robber with a monotone strategy. However, so far there is no upper bound for the monotonicity costs involved. It is conceivable that there is a constant $c \in \mathbb{N}$ such that whenever $n$ cops suffice to catch a robber on a digraph $D$ in any of the two variants, than $c \cdot n$ cops suffice for a monotone strategy. This, however, is left as an open problem.

A different trait we explored in this paper are the limits of an algorithm theory based on directed graph decompositions. We showed that while there are interesting and important examples for natural problems that become tractable on digraphs of small width, many other natural problems remain NP-complete even if the digraphs have very low global connectivity.